\documentclass[11pt]{cernrep}
\usepackage{cite}
\usepackage{here}
\usepackage{amsfonts,amssymb}
\def\nb{}
\def\vec#1{{\bf #1}}
\newif\ifpdf
	\ifx\pdfoutput\undefined
	\pdffalse 
	\else
	\pdfoutput=1 
	\pdftrue
	\fi
\ifpdf
	\usepackage[pdftex]{graphicx}
	\DeclareGraphicsExtensions{.pdf, .jpg}
\else
	\usepackage{graphicx}
		\DeclareGraphicsExtensions{.eps, .jpg}
\fi
\begin{document}
 \title{DOUBLE CHARM PHYSICS}
\author{Jean-Marc Richard%
\footnote{\ e-mail: \texttt{jean-marc.richard@isn.in2p3.fr}}}
\institute{%
Institut des Sciences Nucl\'eaires, Universit\'e Joseph Fourier--CNRS-IN2P3\\
53, avenue des Martyrs, F-38026 Grenoble cedex }
\maketitle
\begin{abstract}
We review the weak-decay and spectroscopy properties of baryons with two charmed quarks. We also present the convergent speculations on exotic mesons $(QQ\bar{q}\bar{q})$ with two heavy quarks and two light antiquarks.
\end{abstract}
\section{INTRODUCTION}
The discovery of the $(b\bar{c})$ ground state \cite{Hagiwara:2002pw}, and that of the $(ccd)$ baryon \cite{Mattson:2002vu,Cooper2002,CooperHere} demonstrates that new sectors of hadron physics are becoming accessible to experiments.

There are several good reasons to study hadrons systems with two $c$ quarks:
\begin{itemize}
\item 
Double charm baryons provide tests of mechanisms proposed to describe the weak decays of charmed mesons and single-charm baryons.
\item
The dynamics of confinement in $(QQq)$ baryons combine the slow relative motion of two heavy quarks with the fast motion of a light quark.
\item
$(QQ\bar{q}\bar{q})$ multiquark states have been predicted, whose stability results from the flavour-independent character of quark forces at short distances and from  pion-exchange between two heavy mesons at large distances.\\
\end{itemize}
\vskip -4pt
These aspects will be reviewed in the next sections. Details will be skipped. Many references will be provided for further reading.
\section{WEAK DECAY OF CHARM}
\subsection{General considerations}
\begin{figure}[ht]
\centering{\includegraphics{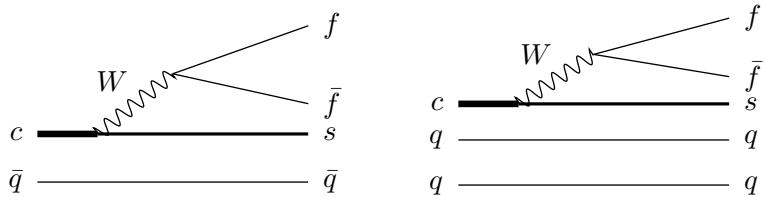}}
\vskip -1cm
\caption{Spectator diagram, for charmed mesons (left) and single-charm baryons (right).\label{Fig:spec}}
\end{figure}

It was a surprise in our community when  the ratio of lifetimes $r=\tau(D^\pm)/\tau(D^0)$ was announced to significantly differ from unity. The preliminary value $r\simeq 4$ even amplified  the shock. Still, the stabilised value $r\simeq 2.5$
\cite{Hagiwara:2002pw} is impressive. With only spectator diagrams such as those of Fig.~\ref{Fig:spec}, all lifetimes would be equal (up to minor phase-space effects) and all semileptonic widths comparable. 
$r\neq1$ reveals important non-spectator effects: interferences between a constituent quark or antiquark,  and another coming  from $c$ or $W$ decay; $W$ exchange between $c$ and $d$ or $s$; to a lesser extent, $W$ formation in the $s$-channel. Further refinements such as penguin diagrams might also be included.
\subsection{Charmed mesons}
There are many data on charmed mesons. In particular, the semileptonic widths are comparable \cite{Hagiwara:2002pw}
\begin{equation}
\Gamma_{\rm SL}(D^\pm)\sim \Gamma_{\rm SL}(D^0)\sim\Gamma_{\rm SL}(D_s)
\sim 0.3\,(\mathrm{ps})^{-1}~.
\end{equation}
There is no major interference effect. So the mechanism of Fig.~\ref{Fig:spec} (left) with $(f,\bar{f})=(e^+,\nu_e)$ or $(\mu^+,\nu_\mu)$ provides all mesons with a similar semi-leptonic rate.

\begin{figure}[h]
\centering{\includegraphics{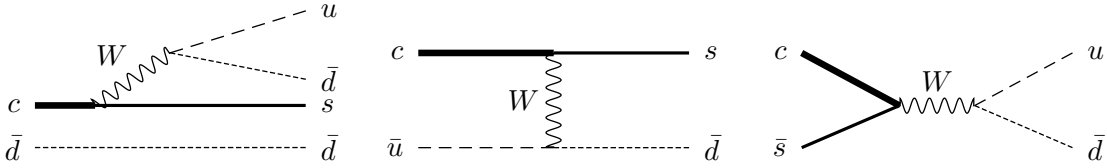}}
\vskip -1cm
\caption{Some mechanisms contributing to differences among the lifetimes of charmed mesons: interferences (left), $W$ exchange (centre), fusion into $W$ (right).\label{Fig:mes-dec}}
\end{figure}

The differences in lifetimes come from the hadronic part. 
The results \cite{Hagiwara:2002pw}
\begin{equation}
\tau(D^0)\sim400\,\mathrm{fs}~,\quad \tau(D_s)\sim500\,\mathrm{fs}~,\quad\tau(D^+)\sim1000\,\mathrm{fs}~,
\end{equation}
indicate that the light antiquark is not a mere spectator.
When $\bar{f}=\bar{d}$ in $W\to f\bar{f}$ decay, this $\bar{d}$ interferes with the $\bar{d}$ of $D^+$. For $D^0$ and $D_s$, a $W$ boson can be exchanged. For $D_s$, there is a small contribution of $W$ formation. Some effects are pictured in Fig.~\ref{Fig:mes-dec}. 

\subsection{Baryons with single charm}
\begin{figure}[h]
\centering{\includegraphics{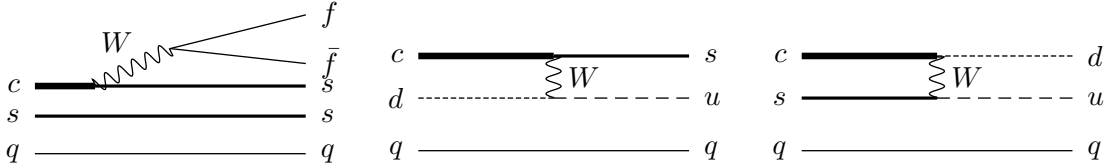}}
\vskip -.2cm
\caption{Some diagrams differentiating the weak decay properties of various single-charm baryons: $ss$ interferences, $W$ exchange for ordinary hadronic decay, $W$ exchange for suppressed decay.\label{Fig:bar-dec}}.
\end{figure}

The above mechanisms have been applied to charmed baryons: $\Lambda_c^+(cud)$, $\Xi_c^+(csu)$, $\Xi_c^0(csd)$ and $\Omega_c(css)$.
A new interference appears with respect to the meson case: the $s$-quark coming from the decaying $c$  might ``feel'' the presence of another $s$. The $W$-exchange contribution receives a larger strength. Annihilation becomes negligible, since requiring an antiquark from the sea. Some typical contributions are shown in Fig.~\ref{Fig:bar-dec}.  The mechanisms can be tested in subclasses of decays, once the statistics becomes sufficient for such filtering. An example is the last diagram of Fig.~\ref{Fig:bar-dec} showing a $W$-exchange contribution to doubly-Cabbibo-suppressed decay.

The main {\em predictions}  \cite{Guberina:1986gd,Fleck:1990ma} are that 
\begin{itemize}
\item there are differences in the semileptonic partial widths $\Gamma_{SL}$, namely
\begin{equation}\label{SL1}
\Gamma_{\rm SL}(\Lambda_c^+)< \Gamma_{\rm SL}(\Xi_c^+)
< \Gamma_{\rm SL}(\Xi_c^0)< \Gamma_{\rm SL}(\Omega_c^0)~,
\end{equation}
\item the lifetimes are ordered as
\begin{equation}\label{Tau1}
\tau(\Omega_c)<\tau(\Xi_c^0)<\tau(\Lambda_c^+)<\tau(\Xi_c^+)~.
\end{equation}
\end{itemize}
Present data do not enable one to check the prediction (\ref{SL1}). The ordering (\ref{Tau1}) of lifetimes is remarkably verified by the data, but the spread of  values seems allways underestimated in theoretical calculations, at least to my knowledge. This is hopefully just a matter of  using more realistic values of some model-dependent parameters, such as the probability to find two quarks at the same location, which enters  the contribution of $W$ exchange.

There are many predictions for exclusive rates, at least for their relative values.
See, e.g., Ref.~\cite{Savage:1990qr} for a flavour of this rich physics. 
\subsection{Weak decays of baryons with double charm}
\begin{figure}[h]
\centering{\includegraphics{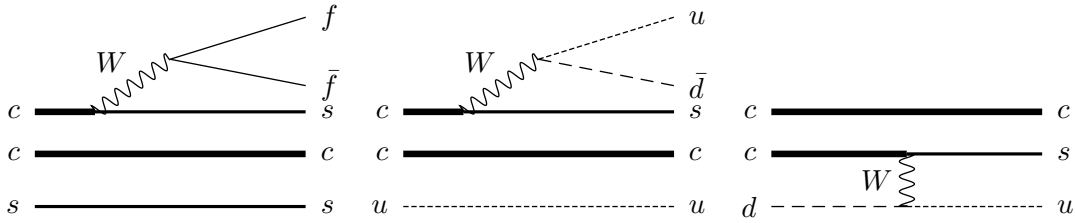}}
\caption{Some diagrams leading to differences among the lifetimes of baryons with double charm.\label{Fig:dbc-dec}}
\end{figure}

\begin{figure}
\begin{center}
\includegraphics[width=.5\textwidth]{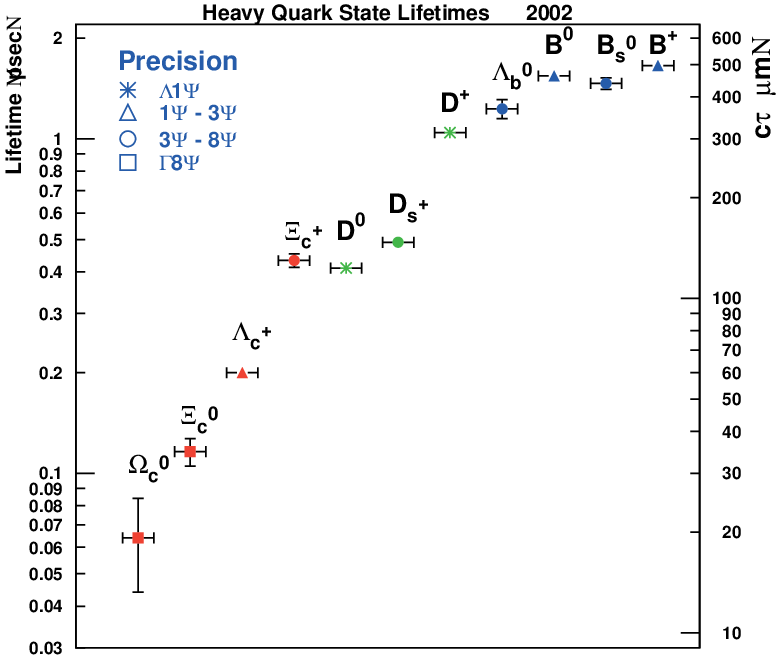}
\caption{Lifetimes of heavy baryons, borrowed from the slides of P.~Cooper at the last  Hyperon Conference \protect\cite{Cooper2002}.}
\label{Fig:lifetimes}
\end{center}
\end{figure}

The same mechanisms have been further applied to baryons with double charm.
Examples are drawn in Fig.~\ref{Fig:dbc-dec}.
There is an overall agreement that the hierarchy of lifetimes is \cite{Fleck:1990ma,Guberina:1999mx}
\begin{equation}\label{Tau2}
\tau(\Xi_{cc}^+)\lesssim \tau(\Omega_{cc}^+)\ll \tau(\Xi_{cc}^{++})~,
\end{equation}
with, perhaps, an underestimate of the magnitude of the effect. For instance, Kiselev et al.\ \cite{Kiselev:1998sy,Kiselev:1999un} predicted $\tau(\Xi_{cc}^+)\sim 400\,$fs, as compared to the
value
$\tau\lesssim 30\,$fs suggested by SELEX data \cite{CooperHere,Mattson:2002vu}. This is one of the reasons leading Kiselev et al.\ \cite{Kiselev:2002an} to cast some doubt about the Fermilab result. See, also, \cite{Yelton:2002hq}. But, again, the lifetime of other charmed baryons was also overestimated by theorists.
In the plot of lifetimes, Fig.~\ref{Fig:lifetimes},  a value as low as $30\,$fs does not look too extravagant an extrapolation.
\section{SPECTROSCOPY OF DOUBLE-CHARM BARYONS}
In the 60's, the flavour group SU(3)$_\mathrm{F}$ has been immediately extrapolated to SU(4) or higher. A better motivation for a fourth quark came from the GIM mechanism \cite{Glashow:1970gm}. At the time where charm was discovered, in the hidden form of $(c\bar{c})$, some classic papers were written on hadrons with charm, including a section on $(ccq)$ states \cite{Gaillard:1975mw,DeRujula:1975ge}.

More detailed studies of $(QQq)$ baryons came in the 80's and later \cite{Fleck:1989mb,Fleck:1990ma,%
Savage:1990di,Savage:1991pr,%
Bagan:1994dy,Roncaglia:1995az,Silvestre-Brac:1996bg,%
Kiselev:2000jb,Tong:1999qs}. No doubt that the recent discovery at SELEX will stimulate further works.

$(QQq)$ baryons are perhaps the most interesting of ordinary hadrons, as they combine in a single bag two extreme regimes:
\begin{enumerate}
\item the  slow relative motion of two heavy quarks, as in charmonium,
\item the fast motion of a light quark. Remember that the electron moves faster in hydrogen than in positronium. Similarly, a light quark is likely more relativistic in heavy-light hadrons than in light mesons.
\end{enumerate}
Hence, $(QQq)$ baryons offer an very interesting laboratory to study confinement.
\subsection{Diquark clustering and excitations}
In the $(QQq)$ wave function, the average $QQ$ separation is smaller than the $Qq$ one. This leads to envisage approximations, such as a quark--diquark picture, to be discussed shortly. The diquark is, however, not frozen. The first excitations arise in the $QQ$ relative motion, i.e., the $(QQq)$ ground state, and its first orbital excitation $(QQq)^*$ are built out of different diquarks.
\subsection{The two-step approximation}
It is rather legitimate to replace the full three-body calculation by a two-step procedure where one
\begin{enumerate}
\item calculates the $QQ$ mass, by solving a two-body problem,
\item calculates the $QQ-q$ mass by solving another two-body problem.
\end{enumerate}
The second step is rather safe. The finite-size corrections are small. For instance, they cancel out exactly for the harmonic oscillator.

As for the first step, one should be aware that the $QQ$ potential is {\em effective}, since it contains both the direct $QQ$ interaction and a contribution from the light quark. For instance, in the harmonic oscillator model, the identity
\begin{equation}
r_{12}^2+r_{23}^2+r_{31}^2={3\over 2} r_{12}^2+ 2\, r_{12-3}^2~,
\end{equation}
demonstrates that 1/3 of the $QQ$ interaction comes from the light quark. Replacing 3/2 by 1 results into an underestimate of energies and spacings by a factor $\sqrt{3/2}$. 
\subsection{The Born--Oppenheimer approximation}
It was used, e.g., by Fleck and Richard \cite{Fleck:1989mb}.
For a given $QQ$ separation $r_{12}$,  the two-centre problem is solved for the light quark, with proper reduced mass. The ground-state energy $E_0(r_{12})$, supplemented by the direct $QQ$ interaction, provides the adiabatic potential $V_{QQ}$. Solving the 2-body problem with this potential gives the first levels. The adiabatic potential built out of the second ``electronic'' energy $E_1(r_{12})$ leads to a second series of levels. This is very similar to the spectroscopy of H$_2^+$ in atomic physics.

Within explicit potential models,  the Born--Oppenheimer approximation can be checked against an accurate solution of the 3-body problem, using for instance a systematic hyperspherical expansion. The approximation is excellent for $(bbq)$ and $(ccq)$, with  $q=u$, $d$ or $s$, or even for $(ssu)$ or $(ssd)$.
\subsection{Typical results}
 In Ref.~\cite{Fleck:1989mb}, $(ccq)$ masses were estimated from a specific variant of the bag model, already used for charmed mesons. The results  turn out to be rather sensitive to details such as centre-of-mass corrections, value of the bag constant, etc. Other bag-model calculations have been performed \cite{Ponce:1979gk}.

Potential models, on the other hand,  tend to give very stable results, when the parameters are varied while maintaining a reasonable fit of lighter hadrons. Typically
\begin{itemize}
\item a ground-state near or slightly above $3.6\,$GeV for the $(ccu)$ or $(ccd)$ ground state,
\item
a hyperfine splitting of about $80\,$MeV between the spin 3/2 and spin 1/2 states,
\item
the first orbital excitation about $300\,$MeV above the ground-state,
\item
the first $(ccs)$ state near $3.7\,$GeV \\
\end{itemize}
Note that models tuned to $(cqq)$ or lighter baryons might underestimate the short-range $QQ$ attraction. If models are adjusted to $(c\bar{c})$ spectroscopy, there is an ambiguity on how to translate it to $cc$. The usual recipe stating that
\begin{equation}
V_{QQ}={1\over2}V_{Q\overline{Q}}~,
\end{equation}
implies pairwise forces mediated by colour-octet exchanges. Small, non-confining, colour-singlet exchanges, as well as three-body forces might complicate the issue.
\subsection{Towards better estimates}
Most existing calculations are of rather exploratory nature, since made when double charm was considered as science fiction, or far future. Meanwhile, the art of QCD has made significant progress.

One could retain from simple potential models that the Born--Oppenheimer approximation provides an adequate framework. The effective $QQ$ potential could be estimated from relativistic models or from lattice calculations, similar to those of the $Q\overline{Q}$ potential or the effective $QQ$ potential in exotic $(QQ\bar{q}\bar{q})$ mesons, on which more shortly. It is hoped that the new  experimental results will stimulate such calculations.

The literature already contains approaches somewhat more ambitious than simple bag or potential models: QCD sum rules \cite{Bagan:1994dy}, string picture \cite{Gershtein:1998sx,Gershtein:2000nx}, etc.
\section{EXOTIC MESONS WITH DOUBLE CHARM?}
\subsection{Minireview on advertised exotics}
The famous $H$ dibaryon proposed by Jaffe \cite{Jaffe:1977yi}, and the less notorious pentaquark $P$ proposed independently by Lipkin \cite{Lipkin:1987sk} and the Grenoble group  \cite{Gignoux:1987cn}, owe their tentative stability to \textit{chromomagnetic} forces. Other mechanisms might lead to stable multiquarks: \textit{chromoelectric} forces and long-range \textit{Yukawa} interaction. These mechanisms have been first proposed 
with  crude approximations for the overall dynamics. It is important to examine to which extent multiquark binding survives all refinements brought in model calculations.

\subsubsection{Hexaquark}
The chromomagnetic interaction \cite{DeRujula:1975ge}
\begin{equation}\label{eq:chromo}
H_{\mathrm cm}=-C\sum_{i<j}{\vec{\sigma}_i .\vec{\sigma}_j\,\tilde{\lambda}_i.\tilde{\lambda}_j\over m_i m_j}\delta^{(3)}(\vec{r}_{ij})~,
\end{equation}
or its bag model analogue \cite{Chodos:1974je}, successfully describes the observed hyperfine splittings such as $\Delta-N$ or $J/\Psi-\eta_c$. The astute observation by Jaffe \cite{Jaffe:1977yi} is that this operator provides a binding 
\begin{equation}\label{eq:H}
(ssuudd)-2(sud)\sim-150\,\mathrm{MeV}
\end{equation}
to the $H=(ssuudd)$ dibaryon with spin and isospin $J=I=0$. This estimates, however, relies on: \\
\begin{enumerate}
\item SU(3)$_\mathrm{F}$ flavour symmetry
\item $\langle \delta^{(3)}(\vec{r}_{ij}) \rangle$ independent of $(i,j)$ pair and borrowed from the wave function of ordinary baryons.\\
\end{enumerate}
Relaxing these hypotheses, and introducing kinetic energy and spin-independent forces in the 6-body Hamiltonian usually spoils the stability of $H$ \cite{Rosner:1986yh,Karl:1987cg,Fleck:1989ff}.
The existence of $H$ is nowadays controversial. It has been searched in many experiments, without success so far. For instance, the doubly-strange hypernucleus
${}_\Lambda^{\phantom{6}}{\!}_\Lambda^6\!\mathrm{He}$ is not observed to decay into $H+\alpha$ \cite{Gal:2002pn}.
\subsubsection{Pentaquark}
If  the calculation made for the $H$ is repeated in the limit where $m(Q)\to\infty$, the same binding
\begin{equation}\label{eq:P}
(\overline{Q}qqqq)-(\overline{Q}q)-(qqq)\sim-150\,\mathrm{MeV}
\end{equation}
is obtained for the pentaquark $(\overline{Q}qqqq)$, $qqqq$ being in a SU(3)$_\mathrm{F}$ triplet \cite{Lipkin:1987sk,Gignoux:1987cn}. All corrections, again, tend to weaken this binding \cite{Karl:1987uf,Fleck:1989ff} so it is not completely sure that the actual pentaquark is stable. See, also, \cite{Leandri:1989su}.

For the case where the  chromomagnetic term (\ref{eq:chromo}) is replaced by Goldstone-boson exchange, see, e.g., the review by Stancu \cite{Stancu:1999xc} and references therein.
\subsection{Tetraquark}
Twenty years ago, it was pointed out that current confining potentials bind $(QQ\bar{q}\bar{q})$ below its dissociation threshold into $(Q\bar{q})+(Q\bar{q})$, provided the mass ratio $m(Q)/m(q)$ is large enough \cite{Ader:1982db}. This \textit{chromoelectric} binding was studied by several authors, in the context of flavour-independent potentials \cite{Heller:1985cb,Heller:1987bt,Carlson:1988hh,%
Zouzou:1986qh,Lipkin:1986dw,Brink:1994ic,Brink:1998as,%
Janc:2000vq,Gelman:2002}  %
 or with lattice QCD \cite{Mihaly:1997ue,Foster:1998wu} (see, also, \cite{Green:1998mv,Green:1998nt}), with a remarkable convergence towards the same conclusion. This somewhat contrasts with the confusion in other sectors of multiquark spectroscopy.
\subsubsection{Favourable symmetry breaking}
Let us consider the limit of a purely flavour-independent potential $V$ for $(QQ\bar{q}\bar{q})$. The situation becomes similar to that of exotic four-body molecules $(M^+,M^+,m^-,m^-)$, which all use the very same Coulomb potential. The hydrogen molecule with $M\gg m$ is much more stable than the positronium molecule Ps$_2$ with $M=m$. If one decomposes the 4-body Hamiltonian as
\begin{equation}
{\cal H}_4=\left[{M^{-1}+m^{-1}\over4}
\left(\vec{p}_1^2+\vec{p}_2^2+\vec{p}_3^2+\vec{p}_4^2\right)+V\right]+
{M^{-1}-m^{-1}\over4}\left(\vec{p}_1^2+\vec{p}_2^2-\vec{p}_3^2-\vec{p}_4^2\right)~,
\end{equation}
the first term, even under charge conjugation, corresponds to a rescaled equal-mass system with \textit{the same threshold} as ${\cal H}_4$. The second term, which breaks charge conjugation, improves the energy of ${\cal H}_4$ (one can applies the variational principle to ${\cal H}_4$ using the symmetric ground state of the first term as a trial wave function).
In the molecular case,  the second term changes the marginally bound Ps$_2$ (or rescaled copy) into the deeply bound H$_2$. In quark models,  an unbound $(qq\bar{q}\bar{q})$ becomes a stable $(QQ\bar{q}\bar{q})$.

The effective $QQ$ potential has been estimated by Rosina et al. \cite{Janc:2000vq} in the framework of empirical potential models, and by Mihaly et al.\  \cite{Mihaly:1997ue} and 
Michael et al. (UKQCD) \cite{Foster:1998wu}, who used lattice simulations of QCD.

The question is obviously: is the $c$ quark heavy enough to make $(cc\bar{q}\bar{q})$ bound when $q=u$ or $d$?
At this point, the answer is usually negative,  most authors stating that $b$ is required to bind $(QQ\bar{q}\bar{q})$ below its $(Q\bar{q})+(Q\bar{q})$ threshold.
\subsection{Deuterium-like binding}
There is, however, another mechanism: pion-exchange or, more generally, nuclear-like forces between hadrons containing light quarks or antiquarks.
This effect was studied by several authors, in particular T{\"o}rnqvist 
\cite{Tornqvist:1991ks}, Manohar and Wise \cite{Manohar:1993nd}, and Ericson and Karl \cite{Ericson:1993wy}. In particular a $D$ and $D^*$ can exchange a pion, this inducing an attractive potential. It is weaker than in the nucleon--nucleon case, but what matters for a potential $g V(r)$ to bind, is the product $gm$ of the strength $g$ and reduced mass $m$.

It is found that $(DD^*)$ is close to be bound, while binding is better established for $(BB^*)$. The result depends on how sharply the long-range potential is empirically regularised at short distances.
\subsection{Combining long- and short-range forces}
A lattice calculation such as those of Refs.~\cite{Mihaly:1997ue,Foster:1998wu} contains in principle all effects. In practice, space is truncated, so long-range forces are perhaps not entirely included. Explicit quark models such as \cite{Janc:2000vq} make specific assumptions about interquark forces, but do not account for pion exchange.

In our opinion, a proper combination of long- and short-range forces should lead to bind $(DD^*)$, since each component is almost sufficient by itself. This is presently under active study.
\subsection{Borromean binding}
There is a further possibility to build exotic, multicharmed systems.
If the interaction between two charmed mesons cannot lead to a bound state (this is presumably the case for $(DD)$, since pion exchange does not contribute here), it is likely that the very same meson--meson interaction binds three or more mesons.
This is known as the phenomenon of ``Borromean'' binding.

For instance, in atomic physics, neither two ${}^3$He atoms nor a ${}^3$He atom and a ${}^4$He atom can form a binary molecule, even at vanishing temperature, but it is found that
${}^3\mathrm{He} {}^3\mathrm{He} {}^4\mathrm{He}$ is bound \cite{Bressanini2002}. Similarly, in nuclear physics, the isotope ${}^6$He is stable against  evaporating two neutrons, or any other dissociation process, while ${}^5$He is unstable. In a	3-body picture, this means that $(\alpha,n,n)$ is stable, while neither $(\alpha,n)$ nor $(n,n)$ have a stable bound state. In short, binding three constituents is easier than two.

\section{Conclusions and outlook}
The results by SELEX \cite{Mattson:2002vu} and BELLE \cite{Abe:2002rb} groups show that we are know able to produce and identify two units of charm in hadron or electron collisions. 

Double charm opens unique perspectives for studying new aspects of weak decays and confining forces, and for producing heavy exotic states.

A step further is triple-charm. The $\Omega_{ccc}$ family was named by Bjorken \cite{Bjorken:1985ei} the ``ultimate goal of baryon spectroscopy''. It will reveals a ``pure'' baryon spectrum, without light quark complications. Comparing $(c\bar{c})$ and $(ccc)$ ordering and spacing pattern will be crucial to check current ideas on the gluon strings picture leading to linear confinement.

\section*{Acknowledgments}
I would like to thank my collaborators on the spectroscopy of charmed particles, in particular J.-P.~Ader, S.~Fleck, M.~Genovese, A.~Martin, S.~Pepin, B.~Silvestre-Brac, Fl.~Stancu, P.~Taxil and S.~Zouzou. I am also grateful to P.~Fayet and X.Y.~Pham for informative discussions on weak decays.
\small

\normalsize
\end{document}